\documentstyle[aps,titlepage]{revtex}

\begin{document}

\title{Bogoliubov Condensation of Gluons and Spontaneous\\
       Gauge Symmetry Breaking in QCD }

\author{V.N. Pervushin, G. R\"opke \thanks{
Permanent address: MPG Arbeitsgruppe  ''Theoretische Vielteilchenphysik'',
         Universit\"at Rostock, D-18051 Rostock, Germany}
and  M.K. Volkov}
\address{Bogoliubov  Laboratory of Theoretical Physics,\\
        Joint Institute for Nuclear Research, 141980, Dubna, Russia}
\author{D. Blaschke and H.-P. Pavel}
\address{MPG Arbeitsgruppe  ''Theoretische Vielteilchenphysik''\\
         Universit\"at Rostock, D-18051 Rostock, Germany}
\author{A. Litvin}
\address{Laboratory of Computing Techniques and Automation\\
Joint Institute for Nuclear Research, 141980 Dubna, Russia}

\maketitle

\vspace{1cm}

\noindent MPG--VT--UR 60/95\\
\noindent December 1995

\begin{abstract}

The problem of the gluonic quasiparticle excitations in QCD is considered
under the
aspect of the condensation of gluon pairs in the ''squeezed'' vacuum.
The present approach is a field theoretical generalization of the
Bogoliubov model which successfully
reproduced the Landau spectrum in the microscopic theory of superfluidity.
We construct a gauge invariant QCD Hamiltonian by formally solving
the Gauss equation such that the physical variables are separated by a
non-Abelian projection operator, instead of fixing a gauge.
By using Dirac quantization
we show that the Bogoliubov condensation of gluon pairs destroys this
projection operator, and the spontaneous appearance of a gluon mass is
accompanied by a longitudinal component for the gluon field in correspondence
with the relativistic covariance. Gauge symmetry is broken spontaneously
since the gauge invariance of the Hamiltonian is not shared by the
vacuum. The squeezed vacuum in the present model is characterized by one
free parameter related to the contraction of a pair of
zero momentum gluon fields
which is fixed from the difference of the $\eta'$ and
the $\eta$ - meson masses ($ U(1)$-problem) and results in a value for
the gluon condensate which is in good agreement with the value obtained
by Shifman, Vainshtein and Zakharov.
\vspace{5mm}
\\
\noindent PACS number(s):  11.15.Ex, 12.38.Aw, 14.70.Dj

\end{abstract}

\section{Introduction}
The problem of the gluon vacuum in QCD has a long history
\cite{savvidy,coleman,leutwyler,belavin,zakharov,shuryak}.
It is well known that at low energies  perturbation theory  does not apply
since the coupling constant becomes large and
the perturbative vacuum becomes unstable due to gluon self-interactions
\cite{savvidy}.
These self-interactions  can lead to a 'reconstruction' of the vacuum and to
the appearance of a condensate \cite{nn,umt}. The presence of the condensate
is important for the physical properties of the low energy sector of QCD.
Of particular interest is the modification of the quasiparticle spectrum and
the occurence of massive collective excitations to be considered in this work.

In the literature two different kinds
of such condensates have been considered,
the {\it coherent} and the {\it squeezed} one.
In the coherent condensate gluon field excitations are
found by a {\it transitive} transformation, i.e.
shifting the fields to the solution of the classical
equations \cite{savvidy,coleman,leutwyler}
(e.g. instantons \cite{belavin,zakharov,shuryak}).
It is characterized by the condensation of single gluons and thus by
a nonvanishing vacuum expectation value of the gluon field $A$:
\begin{equation}
<A>\ \ \neq\ \ 0~.
\end{equation}
In the squeezed condensate on the other hand the gluon states are
constructed by a {\it multiplicative} Bogoliubov
transformation of the gluon fields \cite{celenza,biro1,biro2,mishra1,mishra2}.
The squeezed vacuum is characterized by the condensation of colourless,
scalar gluon pairs and could thus be realized through
\begin{equation}
<A>\ \ =\ \ 0\ ,\ \ <A^2>\ \ \neq 0\ .
\end{equation}

There has been a lot of activity to construct
a stable coherent vacuum in the gluon sector of QCD
\cite{savvidy,coleman,leutwyler}.
The problem in this case is that there are no stable
quasiclassical solutions to the
Yang-Mills equations in Minkowski space \cite{coleman,leutwyler}.
In recent years
the squeezed condensate (called here {\it Bogoliubov condensate})
has become a topic of great interest, see e.g.
\cite{celenza,biro1,biro2,mishra1,mishra2}.
Its investigation for non-Abelian fields faces the following problems:
\begin{enumerate}
\item
One has to find the adequate degrees of freedom to construct the gluon
condensation in a squeezed vacuum.
\item
Since the squeezed vacuum is most naturally described in the Hamiltonian
formalism one has to extract the gauge invariant oscillator - like
field variables from the non-Abelian QCD action. Recall that by using different
gauges for instance
Biro \cite{biro1,biro2} and Mishra \cite{mishra1,mishra2} get different
results for the condensate.
\item
The condensate leads to spontaneous gauge symmetry breaking with the
appearance of massive gluons. This has to
be accompanied by the corresponding
generation of the longitudinal components for the massive gluon fields.
Recent papers on ''squeezed'' gluon states
\cite{biro1,biro2,mishra1,mishra2} indeed obtain a constituent gluon mass.
However, since they fix the gauge, their gluon fields do not have a
longitudinal
component such that they are unable to provide the number of degrees of
freedom required for the description of a massive gluon vector field
\cite{slavnov}.

\end{enumerate}

With respect to the last topic, we mention that
mass generation for gauge fields via spontaneous gauge symmetry breaking
is a general problem familiar also from
other gauge field theories.
In the unified theory of electroweak interactions a
consistent description of massive vector fields can be given
due to the presence of a scalar Higgs field in the Lagrangian
which generates the longitudinal component of the massive gauge bosons
W and Z. The very interesting alternative, that the gauge bosons obtain
their mass by spontaneous gauge symmetry due to radiative corrections,
without introduction of an external Higgs field,
has been proposed by Coleman and Weinberg \cite{cw}.
This possibility is of great importance since for SU(N) gauge theories
the introduction of an external
Higgs field does not lead to a mass term for the gluons, as was shown by
Georgi and Glashow \cite{Georgi}.
The concept of spontaneous gauge symmetry breaking by radiative corrections
is therefore very attractive for the case of QCD and has been followed up
to now \cite{biro1,biro2,mishra1,mishra2}.

In the present work we study the possibility of gluon condensation
in a squeezed vacuum in view of the Bogoliubov model
\cite{nn} of the weakly nonideal Bose gas. In particular we investigate
the influence of a squeezed condensate on the gluon quasiparticle spectrum
in the low-energy region of QCD.
Note that the Bogoliubov theory was the first to  explain the experimentally
observable spectrum of collective excitations of superfluid $^4$He.
This spectrum cannot be obtained by resummations of the conventional
perturbation theory series.
As Bogoliubov has shown, the collective
exitations are determined by the ''condensate'' of particles with zero
momentum and finite density. A first connection between Bogoliubov
condensation and the squeezed vacuum in field theory
(massless $\lambda \phi^4$)
was made by Castorina and Consoli \cite{CC}, see also \cite{PaBla}.
An application of the
concept of the Bogoliubov model to QCD,
however, has to our knowledge not been carried out by now.

In a first attempt to generalize the Bogoliubov model to QCD
we use the infrared
singularity of massless theories to squeeze the zero momentum mode
which leads to massive gluonic quasiparticles
in the nonzero momentum sector.
The free squeezing parameter is fixed from the $\eta-\eta'$ mass difference.
Our coresponding value for the gluon condensate is in
reasonable agreement with that obtained by Shifman, Vainshtein and Zakharov
\cite{zakharov} which supports our semi-phenomenological approach.

Concerning the above mentioned problems (2) and (3),
we try to solve the problem of
the appearance of a constituent gluon mass  using
a gauge invariant scheme for the elimination of the unphysical components
of the gluon vector field \cite{dirac,perv1,perv2,fj,kp} which does not
require the gauge-fixing as  initial supposition.
This scheme is based on
the construction of   projection operators by formally solving the Gauss law
constraint. We show that these projectors are destroyed by
the interaction of gluons with the squeezed vacuum. As result
a constituent gluon mass appears together with the necessary
longitudinal components.
This is the central result of our paper and is quite
in analogy to spontaneous chiral symmetry breaking
in the quark sector \cite{ad,lopr,pk2s}. There, the appearance of
constituent quark masses due to the interaction of quarks with the squeezed
vacuum is accompanied by the destruction of the chiral projection
operator which leads to the necessary
increase of the number of spinor field components from two (Weyl spinors)
to four (Dirac spinors).

The paper is organized as follows:
In Section \ref{sec:field}, the Bogoliubov model
of a weakly nonideal Bose gas is generalized to field theory.
We give a field
theoretical description of the condensation phenomenon by the use of the
squeezed vacuum and discuss the conventional local
$\lambda \phi^4$ theory. The relation to  the Bogoliubov model
for the weakly nonideal Bose gas is given in Appendix \ref{app:bogmod}.
In Section \ref{sec:qcd}
the homogeneous colourless
Bogoliubov condensate of gluons is introduced in QCD
where the unphysical degrees of freedom are eliminated by applying
projection operators instead of fixing a gauge.
We also discuss spontaneous gauge symmetry breaking and
the corresponding occurrence of a massive gluon quasiparticle spectrum.
In Section \ref{sec:app}  the squeezing parameter is fixed
>from the $\eta' - \eta$ mass difference.
In Section \ref{sec:end} we present the conclusions.

\section{Bogoliubov condensation in quantum field  theory}
\label{sec:field}

In order to introduce some notations and methods needed for our investigation
of the squeezed condensate in the rather complicated QCD,
we first consider massless $\lambda\phi^4$ theory with the Hamiltonian
\begin{equation}
H=\int d^3x[\pi(x)^2+\left(\partial_i\varphi(x)\right)^2+{\lambda\over 4!}
\varphi^4(x)]
\end{equation}
as the simplest example
of an interacting bosonic theory which is renormalizable.
The theory is quantized by turning the classical fields
$\varphi({\bf  x},t),\pi({\bf  x},t)$ to Schr\"odinger operators
$\varphi({\bf  x}),\pi({\bf  x})$
and imposing the canonical commutation relations
\begin{equation}
[\pi({\bf  x}),\varphi({\bf  x}')]=-i\delta({\bf  x}-{\bf  x}')~.
\end{equation}
In the momentum representation defined by
\begin{equation}
\label{FouTra}
\varphi_p={1\over\sqrt{V}}\int d^3x
\mbox{\large e}^{i{\bf  p}{\bf  x}}\varphi({\bf  x}),\ \ \ \
\pi_p={1\over\sqrt{V}}\int d^3x
\mbox{\large e}^{i{\bf  p}{\bf  x}}\pi({\bf  x})~,
\end{equation}
the Hamilton operator is
\begin{equation}
\label{hf1}
:{ H}[\varphi,\pi]:= \frac{1}{2} \sum_p  \left[:\pi_p \pi_{-p}: + p^2
:\varphi_p \varphi_{-p}: \right]
+ \frac{\lambda}{4!V} \sum_{p_1 p_2 p_3 p_4} \delta_{p_1+p_2+p_3+p_4,0}
:\varphi_{p_1}\varphi_{p_2}\varphi_{p_3}\varphi_{p_4}: ,
\end{equation}
with the commutation relations
\begin{equation}
\label{crphipi}
[\pi_p,\varphi_{p'}]= - i\delta_{p,-p'} ~,~~
[\varphi_p,\varphi_{p'}]= [\pi_p,\pi_{p'}]= 0~.
\end{equation}
In (\ref{hf1}) we have introduced normal ordering with respect to the
creation and annihilation operators
$a_p, a_p^+$ defined according to
\begin{eqnarray}
\label{phipi}
\varphi_p =\sqrt{\frac{1}{2  \tilde{\omega}(p)}} (a_p + a^+_{-p})\ , \;\ \ \
\pi_p = i \sqrt{\frac{\tilde{\omega}(p)}{2 }} (-a_p + a^+_{-p})     ~,
\end{eqnarray}
with an arbitrary function $\tilde{\omega}(p)$.
The operators $a_p, a_p^+$ satisfy the commutation relations
\begin{equation}
\label{craa}
[a_p, a_{p'}^+]= \delta_{p, p'}~,~~[a_p, a_{p'}]= [a_p^+, a_{p'}^+]= 0 ~.
\end{equation}
The corresponding vacuum $|0> $ is defined by $a_p |0> = 0$, and the Fock
space is given as
\begin{eqnarray}
\label{fockstate}
\{\,|\,\Phi >\} = \,|\,0 >;\;\;\;  a^+_p |0 > = |\,p >, ... ~.
\end{eqnarray}

We note that the special choice $\tilde{\omega}(p)=|p|$ diagonalizes the
free ($\lambda$ independent) part of the Hamiltonian. For this case,
however, $a_0$ and $a_0^+$ are not defined which corresponds to the well-known
infrared singularity of massless theories.
Generalizing the Bogoliubov model to field theory we should use the infrared
singularity of massless theories to squeeze the zero mode
by populating it macroscopically with massless particles.
Then, we diagonalize the nonzero mode single particle part of the
resulting squeezed Hamiltonian by changing from particles to quasiparticles
whose dispersion relation $\tilde{\omega}(p)$ is finally
determined selfconsistently. For the simple case of $\lambda\phi^4$
this has been carried out in detail \cite{PaBla}. Similar to the Bogoliubov
model this leads to renormalization of the bare parameters like the coupling
constant $\lambda$.

For the time being we leave $\tilde{\omega}(p)$ open and suppose
that the vacuum of the theory (\ref{hf1})
contains a large number of quasiparticles with zero momentum ($p=0$).
We construct this vacuum using the unitary squeezing operator
\begin{equation}
\label{ub}
U_B (\varphi_0,\pi_0) =
\exp \left (i\frac{f_0}{2} (\pi_0 \varphi_{0} + \varphi_0 \pi_{0})
\right )~,
\end{equation}
where $f_0$ is a very large parameter to be fixed later.
The operator $U_B$ transforms the Fock space of states to the Bogoliubov space
of states
\begin{eqnarray}
\label{bogstate}
|\, \Phi_B >\equiv U_B^{-1} \,|\, \Phi > ~.
\end{eqnarray}
In quantum optics these states are called
'squeezed states', see e.g. \cite{optics}.

Applying the unitary transformation (\ref{ub}), we can define
the new field operator $\varphi^B_0$  and its momentum $\pi^B_0$ by means of
\begin{eqnarray}
\label{bt}
\varphi^B_0 = U_B ^{-1}\varphi_0 U_B = {\rm e}^{-f_0} \varphi_0~,\nonumber\\
\\
\pi^B_0 = U_B ^{-1}\pi_0 U_B = {\rm e}^{f_0} \pi_0 \nonumber~,
\end{eqnarray}
which satisfy the same algebra of commutation relations
as the initial ones (\ref{crphipi}).

We shall now carry out the squeezing of the zero mode part of the Hamiltonian
by applying a Wick reordering procedure to the Bogoliubov vacuum $|0_B>$.
Note that under the squeezing transformation (\ref{bt}) the contraction
of a pair of field operators is left invariant,
\begin{equation}
C=
<0 \,|\,\varphi_{0}\varphi_{0} \,|\,0> =
<0_B\,|\,\varphi^B_{0} \varphi^B_{0}\,|\,0_B>~.
\end{equation}
The normal ordering
of the Bogoliubov fields $\varphi_B$ with respect to the Bogoliubov vacuum
 (which is denoted as  $::\varphi^B_{0} \varphi^B_{0} ::$)
has the same form as the normal ordering of the original
fields with respect to the Fock vacuum (\ref{fockstate})
\begin{eqnarray}
\label{no}
C = \varphi_{0} \varphi_{0} - :\varphi_{0} \varphi_{0}:
= \varphi^B_{0} \varphi^B_{0} - ::  \varphi^B_{0} \varphi^B_{0} ::~~.
\end{eqnarray}
To reorder the Hamiltonian (\ref{hf1})
with respect to the new vacuum $|0_B >$ we use Eqs.(\ref{bt}) and (\ref{no}).
Reordering of the quadratic term gives
\begin{equation}
\label{bc}
:\varphi_{0} \varphi_{0}: =
::\varphi_{0}^B \varphi_{0}^B::~
\mbox{\large{e}}^{2f_{0}}~ + ~\tilde{C}  ~,
\end{equation}
where
\begin{eqnarray}
\label{bfi}
\tilde{C}=
C
\left (\mbox{\large{e}}^{2f_{0}} - 1 \right )~.
\end{eqnarray}
Analogously we have
\begin{equation}
:\pi_{0} \pi_{0}: = ::\pi_{0}^B \pi_{0}^B::~
\mbox{\large{e}}^{-2f_{0}}~+~ C^{\pi}
\left (\mbox{\large{e}}^{-2f_{0}} - 1 \right ),
\end{equation}
with $ C^{\pi}=$
Reordering of the quartic term gives
\begin{eqnarray}
:\varphi_{0}\varphi_{0}\varphi_{0}\varphi_{0}: &=&
::\varphi_{0}^B \varphi_{0}^B \varphi_{0}^B \varphi_{0}^B::
\mbox{\large{e}}^{4f_{0} }
+:: \varphi_{0}^B \varphi_{0}^B::
\mbox{\large{e}}^{2f_{0}}
\tilde{C}  +(5\,\, {\rm permutations})\nonumber\\
& & +\tilde{C}^2 + (2 \,\,{\rm permutations})~.
\end{eqnarray}
For applications in QCD we quote here also the general Wick reordering
result for any polynomial $F(\varphi_0)$
\begin{equation}
:F(\varphi_0):=\exp\{{1\over 2}\tilde{C}
{d^2\over d b^2}\}::F(\varphi^B_0\mbox{\large{e}}
^{f_0}+b)::\Big|_{b=0}~.
\end{equation}
As result of the reordering of $H=H_0+H'$,
where $H'$ is the Hamiltonian of nonzero momentum excitations ($p\neq 0$), we
obtain
\begin{eqnarray}
:H:&=& ::H_0(\phi_0):: + :H':~,\nonumber\\
\label{hfb}
:H'[\varphi,\pi]: &=&
E_0
+ :H^{(2)}[\varphi,\pi]:
+ :H^{(4)}[\varphi]:~,
\end{eqnarray}
where
\begin{eqnarray}
\label{h0}
E_0&=& 3{\lambda\over 4!V}\tilde{C}^2 + C^{\pi}\\
:{H}^{(2)}[\varphi,\pi]: &=& \frac{1}{2}\sum_{p\neq 0}
\left\{:\pi_{p} \pi_{-p}:  +
\left[p^2 +
{\lambda\over 2V} \tilde{C} _{00}\right]
:\varphi_{p} \varphi_{-p}: \right\} ~,\\
:{ H}^{(4)}[\varphi]:  &=&{\lambda\over 4!V}
\sum_{p_1,p_2,p_3,p_4\neq 0}\delta_{p_1+p_2+p_3+p_4,0}
:\varphi_{p_1}\varphi_{p_2}\varphi_{p_3}\varphi_{p_4} :~.
\end{eqnarray}
The zero momentum operator $::H_0(\phi_0)$ containing terms proportional to
$::\varphi_{0}^B \varphi_{0}^B::$ and
$::\varphi_{0}^B \varphi_{0}^B \varphi_{0}^B \varphi_{0}^B::$
describes excitations of the condensate and has not been written explicitely
here.
Note that for very large $f_0$ the second term in the expression (\ref{h0}) for
$E_0$ is much smaller than the first  one (cf. (\ref{bfi})) and can be
neglected. Hence we find a condensate energy density
\begin{equation}
\label{Condene}
\epsilon_0\equiv{E_0\over V}=
{\lambda\over 8}\left({\tilde{C}\over V}\right)^2
\end{equation}
and a bosonic quasiparticle mass $m_B$,
\begin{equation}
\label{Bogmass}
m_B^2 \equiv  \lambda {\tilde{C}\over 2V}~,
\end{equation}
appears.
Diagonality of the one-particle part $:H^{(2)}:$ of the reordered
Hamiltonian (\ref{hfb}) demands that we put the quasiparticle energy
$\tilde{\omega}(p)$ in Eq. (\ref{phipi}) to
\begin{equation}
\tilde{\omega}(p)=\sqrt{p^2+m_B^2}~.
\end{equation}
 The effective Hamiltonian depends (through $\epsilon_0$ and $m_B$) on
the free parameter $\tilde{C}$.

Using Eq. (\ref{Bogmass}) we can eliminate the parameter $\tilde{C}$
>from the expression for the condensate energy density $\epsilon_0$ in Eq.
(\ref{Condene}) to obtain
\begin{equation}
\epsilon_0 =\frac{m_B^4 }{2 \lambda}~,
\end{equation}
which has the same nonanalytic dependence on the coupling constant as
that of the Higgs mechanism of mass generation.
Obviously, the Bogoliubov mechanism of spontaneous mass generation
presented above differs from the Higgs mechanism by the representations
of the vacuum and the interaction of the quasiparticles.
The Higgs mechanism corresponds to the coherent vacuum representation, while
the Bogoliubov one - to the squeezed vacuum, see Appendix \ref{app:bogmod}.
The introduction of the Bogoliubov condensate is related to the Wick
reordering procedure with respect to a new Fock space.


\section{Bogoliubov condensate in QCD}
\label{sec:qcd}

After the introductory generalization of Bogoliubov condensation for
superfluid $^4$He to
massless $\lambda\phi^4$ theory above it is attractive
to suppose that also the gluon vacuum of QCD can be considered as a
homogeneous colourless condensate of gluon pairs.
First steps in this direction where undertaken in Celenza and Shakin
\cite{celenza}.
The corresponding treatment in QCD is far more complicated than in
$\lambda\phi^4$ due to the
fact that QCD is a gauge theory with unphysical degrees of
freedom in the Lagrangian which have to be eliminated before quantization.
According to Dirac \cite{dirac}, only the spatial components of the gauge
fields are dynamical and have to be quantized. The time components obey
constraint equations (Gauss laws) and have to be eliminated.

As in the simpler case of massless $\lambda\phi^4$ theory the
squeezed condensate
is described by the procedure of Wick reordering with a free parameter
$\tilde{C}$ and leads both to a vacuum energy and to a mass term for the field.
In QCD, however, the presence of a condensate and the corresponding generation
of a mass term for the gauge field leads to spontaneous gauge symmetry
breaking, since the gauge invariance of the Hamiltonian is not shared by
the vacuum. The problem of
the appearance of a constituent gluon mass und the corresponding
resurrection of the longitudinal component of the massive quasigluons is
solved by using
a projection scheme for the elimination of the unphysical components
of the gluon vector field \cite{dirac,perv1,perv2,fj,kp} instead of
gauge-fixing.
The projectors are obtained by formally solving the Gauss law
constraint and appear in the kinetic energy term of a reduced gauge invariant
QCD Hamiltonian.
We show that these projectors are destroyed by
the interaction of gluons with the squeezed vacuum so that
the constituent gluon mass appears together with the necessary
longitudinal components.
The presence of a squeezed condensate leads to spontaneous gauge symmetry
breaking:
The gauge invariance of the QCD Hamiltonian is not shared by
the vacuum.

Finally we fix the free parameter $\tilde{C}$ of our squeezed vacuum
by estimating a value for the quasigluon mass from the $\eta'-\eta$
mass difference and comparing the corresponding condensate energy density
to the well known value obtained by Shifman et al. \cite{zakharov}.
We shall see that they are in good agreement.

\subsection{QCD Hamiltonian and Gauss law}

We start from the QCD Lagrangian
\begin{eqnarray}
\label{1}
{\cal  L}(A) = -\frac{1}{4} F_{\mu\nu}^a F^{\mu\nu a}~,
\end{eqnarray}
where $F_{\mu\nu}^a$ is the field strength tensor
\begin{eqnarray}
F^a_{\mu\nu} = \partial_{\mu} A^a_{\nu} -
\partial_{\nu} A^a_{\mu} + g f^{abc} A^b_{\mu} A^c_{\nu}~.
\end{eqnarray}
In the following, we use the notation
\begin{eqnarray}
{A}_{\mu} =  g \frac{A^a_{\mu} \lambda^a}{2i}~,
\end{eqnarray}
where $ g$ is the coupling constant.
Due to the gauge invariance
\begin{eqnarray}
\label{GI}
{A}^v_{\mu} = v({A}_{\mu} + \partial_{\mu})v^{-1}~,
\end{eqnarray}
\begin{eqnarray}
{\cal  L} (A^v) = {\cal  L} (A)~,
\end{eqnarray}
this classical Lagrangian contains only $3(N_c^2-1)$ degrees of
freedom instead of the  $4(N_c^2-1)$ components of the ${A}$ field
($N_c$ is the number of colours).
For the construction of the Hamiltonian one usually introduces chromoelectric
and chromomagnetic fields
\begin{eqnarray}
E^a_i&=&\dot{A}^a_i - D^{ab}_i({\bf  A}) A_0^b~,\\
B^a_i({\bf  A}) &=& \frac{1}{2}~\epsilon_{ijk}
F_{jk}^a = \epsilon_{ijk}D^{ab}_j({\bf  A})
A^b_k~,
\end{eqnarray}
with
$\dot{A}_i = {\partial_0} A_i $ and
the covariant derivative
\begin{eqnarray}
D^{ab}_i ({\bf  A}) = \delta^{ab} \partial_i +  g~{f^{acb}} A^c_i~.
\end{eqnarray}
The magnetic field satisfies the Bianchi  identity
\begin{eqnarray}
\label{BI}
D^{ab}_i ({\bf  A}) B^b_i({\bf  A}) \equiv 0 ~,
\end{eqnarray}
which can be interpreted as a generalized transversality  of the magnetic
field.

In order to construct the Hamiltonian, we have to find the canonical
momenta. We see that the Lagrangian (\ref{1}) does not contain time
derivatives of the zero components of the gluon fields. The corresponding
Euler-Lagrange equations are therefore constraints
(the Gauss laws):
\begin{eqnarray}
\label{Gaulaw}
D^2_{ab} ({\bf  A}) A^b_0 = D^{ab}_i ({\bf  A}) \dot{A}^b_i~,
\end{eqnarray}
where $D^2_{ab} = D^{ac}_i D^{cb}_i$.
In terms of the electric field the Gauss laws (\ref{Gaulaw}) read
\begin{equation}
\label{clGau}
G^a({\bf  A},{\bf  E})\equiv D^{ab}_i({\bf  A})E^b_i=0~.
\end{equation}
The canonical momenta to the spatial fields $A_i^a$ are
the electric fields:
\begin{equation}
{\delta{\cal  L}\over \delta \dot{A}^a_i}=E_i^a ~,~~i=1,2,3~.
\end{equation}
The Hamiltonian can now be written as
\begin{equation}
\label{clHam}
H({\bf  A},{\bf  E})=\int d^3x {1\over 2}\left[{E_i^a}^2+{B_i^a}^2\right]~.
\end{equation}
In the classical theory we thus have the Hamiltonian (\ref{clHam}) together
with the Gauss constraint (\ref{clGau}).

\subsection{Quantization}

In order to quantize the theory one could write the Hamiltonian
completely in terms of gauge invariant variables and their canonical
conjugate momenta and then impose the canonical commutation relations
only on these gauge invariant variables. Different to the case of QED,
this leads to inconsistencies in QCD, as shown in detail in Appendix B.

A more successful alternative way is Dirac quantization \cite{dirac},
where one imposes the
canonical commutation relations on the original $A_i$ and $E_i$:
\begin{equation}
\label{QCDCCR}
[E_i^a({\bf  x}),A_j^b({\bf  x}')]=
i\delta^{ab}\delta_{ij}\delta({\bf  x}-{\bf  x}')~.
\end{equation}
Both the Hamiltonian $H$ in (\ref{clHam}) and the Gauss function $G^a$ in
(\ref{clGau})
then become operators
satisfying
\begin{eqnarray}
\label{GIC}
[H,G^{a}({\bf  x})] &=& 0~, \\ \quad
[ G^{a}({\bf x}),G^{b}({\bf x'})]
 &=& if^{abc}G^c({\bf  x})\delta({\bf  x}-{\bf  x}').
\end{eqnarray}
Since $G^a$ can be interpreted as the infinitesimal generator for gauge
transformations these two commutation relations express the gauge invariance
of the Hamiltonian (under small gauge transformations).
Note that the Hamilton operator obtained from the classical
Hamiltonian (\ref{clHam}) with the cartesian fields $E_i$ and $A_i$ has the
correct operator ordering \cite{ChrLee}.
As pointed out by Jackiw \cite{RJ},
the Gauss law $G^a=0$ cannot be taken as an operator equation
since it would lead to inconsistency with the Dirac commutation relations
(\ref{QCDCCR}).
Jackiw then suggested that the Gauss law should be implemented
by demanding that a physical state satisfies the Schr\"odinger equation
(with energy eigenvalue $\cal E$) and is annihilated by the Gauss law operator
\begin{eqnarray}
H({\bf  A},{\bf  E})|\Phi>&=&{\cal E} |\Phi>~,\\
\label{GauCon} G^a({\bf  A},{\bf  E})|\Phi>&=& 0~.
\end{eqnarray}
The second equation is the condition of gauge invariance of
the physical states.
The Gauss law constraint (\ref{GauCon}) can then at least in principle
be implemented by use of unitary transformations \cite{GJ} and
is still under lively discussion \cite{Lenz}.
The resulting kinetic term in the Hamiltonian is very complicated.

The requirement of gauge invariance of the physical states
expressed by (\ref{GauCon}), however, is too
restrictive. It does not allow for the possibility of spontaneous
breaking of the gauge symmetry in analogy to spontaneous symmetry
breaking in the Higgs model and spontaneous chiral symmetry breaking.
If the gauge symmetry is broken spontaneously, the gauge invariance
of the Hamiltonian is not shared by the vacuum.

We can arrive at a gauge invariant Hamiltonian
without demanding gauge invariance of the physical states, especially
of the vacuum
\begin{equation}
G^{a}({\bf  E},{\bf  A})|0_B>\neq 0~,
\end{equation}
by starting from a gauge invariant reduced
classical Hamiltonian. This is achieved by a projection method described in
the following.

Using the formal solution of the Gauss equations (\ref{Gaulaw})
\begin{equation}
\label{10}
A^a_0[{\bf  A}] = \frac{1}{D^2_{ab}({\bf  A})} D^{bc}_i ({\bf  A})\dot{A}_i^c~,
\end{equation}
the electric field can be written as
\begin{equation}
\label{transvE}
E^a_i=\Pi_{ij}^{ab}({\bf  A})\dot{A}_j^b~,
\end{equation}
with the projection operator
\begin{equation}
\label{pa}
\Pi_{ij}^{ab}({\bf  A}) =
 \delta_{ij}\delta^{ab}-D^{ac}_i({\bf  A}){1\over D^2_{cd}({\bf  A})}
D^{db}_j({\bf  A})~.
\end{equation}
We assume that zero modes of the differential operator $D^2_{ab}(A)$
are absent. Consideration of zero modes is under current investigation
\cite{kp}.
In the case of $A=0$, this projection operator reduces to the transverse one
$\Pi_{ij}^{ab}({\bf  A}=0) =\delta^{ab} \delta^T_{ij}\equiv\delta^{ab}
\left(\delta_{ij}-\partial_i\partial_j/\partial^2\right)$.

The gauge invariant reduced Lagrangian can be written as
\begin{equation}
\label{redLag}
{\cal  L}^{Red}({\bf  A})=
{1\over 2}\left[\left(\Pi_{ij}^{ab}({\bf  A})\dot{A}_j^b\right)^2-
                        {B_i^a}^2({\bf  A})\right]~.
\end{equation}
Note that we still have
\begin{equation}
{\delta{\cal  L}^{Red}\over \delta \dot{A}^a_i}=E_i^a~, ~~i=1,2,3~.
\end{equation}
Due to the property
$\Pi^2=\Pi$ of the projection operator the gauge invariant reduced
Hamiltonian can be written in the form
\begin{eqnarray}
\label{redHam}
H^{Red}({\bf  A},{\bf  E})=
        \int d^3x
 {1\over 2}\left[E_i^{a}\Pi^{ab}_{ij}({\bf  A})E_j^b+{B_i^{a}}^2({\bf  A})
\right]~.
\end{eqnarray}
The non-Abelian projection operator has been inserted between the cartesian
electric fields $E_i$, which as variables of the Hamiltonian lost their
transversality property. The above form
(\ref{redHam}) is only one of many possible forms $E\Pi E$, $(\Pi E)^2$,
$(\Pi E)\Pi (\Pi E)$,... Whereas in QED they are equivalent due to
the property
$\Pi^2=\Pi$ and the possibility to perform partial integrations, in QCD they
are inequivalent due to the presence of the $A$ field in the covariant
derivatives and lead to different operator orderings of $E$ and $A$ after
quantization. The simplest choice $E\Pi E$ in (\ref{redHam})
will be correct at least for our investigation of a squeezed homogeneous
condensate, as discussed in the next paragraph.
Although the form (\ref{redHam}) is gauge invariant classically, we did not yet
succeed in showing explicitly, that the corresponding
Hamilton operator satisfies (\ref{GIC}).

Thus in our treatment the role of gauge fixing is played
by the projection operator (\ref{pa}), for deatils see \cite{perv1,perv2,kp}.
The nonabelian chromomagnetic field projects onto the generalized
transverse component of the $A$ field quite analogous to the  form
$E_i^a \Pi_{ij}^{ab}(A) E_j^b$ for the chromoelectric field.


\subsection{Spectrum of quasigluon excitations}
\label{ssec:qge}

We shall consider the squeezed vacuum containing a colourless homogeneous
condensate of gluon pairs for the Hamiltonian
\begin{equation}
\label{QCDHAM}
:H({\bf  A},{\bf  E}): = \int {\rm d}^3 x ~ \frac{1}{2} \left[
:E_i^a \Pi_{ij}^{ab}({\bf  A}) E_j^b: + :{B^a_i}^2({\bf  A}):
 \right]~.
\end{equation}
We have introduced normal ordering with respect to creation $a_p^+$ and
annihilation operators $a_p$ defined with respect to some open
$\tilde{\omega}(p)$ in close analogy to our definitions (\ref{FouTra}) and
(\ref{phipi}) introduced in Section II for the $\lambda\varphi^4$ model.
A four-gluon interaction term  occurs in  the $AAAA$ term of the magnetic part
${B^{a}_i}^2({\bf  A})$ and in the kinetic term
 $E_i^a \Pi_{ij}^{ab} E_j^b$.
In analogy to the $\lambda \varphi^4$ model we perform Wick reordering
to the new squeezed vacuum and consider a
homogeneous and colourless condensate ($f_{p\neq0}=0, f_0\neq 0$) with the
contraction
\begin{eqnarray}
\label{aa}
<0 \,|\,A_i^a(p_1)A_j^b(p_2) \,|\,0>&=&  <0_B\,|\,(A^B)_i^a(p_1)
(A^B)_j^b(p_2)\,|\,0_B> =
\delta_{ij} \delta^{ab} \delta_{p_1,0}\delta_{p_2,0} C~~,
\end{eqnarray}
where $A({\bf  p}),E({\bf  p})$ are the Fourier transforms of
 $A({\bf  x}),E({\bf  x})$ in analogy
to (\ref{bc}),
and the reordering formula for zero momentum gluon fields
\begin{equation}
\label{apa}
:A_k^a(p=0) A_l^b(p=0): =
::{(A^B)}^a_k(p=0){(A^B)}^b_l(p=0):: {\rm e}^{2f_0} + \tilde{C} \delta^{ab}
\delta_{k,l}~,
\end{equation}
is in analogy to Eq. (\ref{bc}), where
$\tilde{C}= C \left (\mbox{\large{e}}^{2f_{0}} -1 \right)$.
The corresponding reordering formula for the
quartic term is calculated in Appendix C.

The Wick reordering of the magnetic part of the
Hamiltonian (\ref{17}),
\begin{eqnarray}
\frac{1}{2}\int {\rm d}^3 x ~:{B_i^a}^2({\bf  A}): &=&
\frac{1}{2}\int {\rm d}^3 x ~\left\{:(\partial_j A_k^a) \delta^T_{kl}
(\partial_jA_l^a):
+2 g f^{abc} :(\partial_j A_k^a) A_j^b A_k^c:
+\frac{1}{2} g^2 f^{abc} f^{ade} : A_j^bA_k^cA_j^dA_k^e : \right\}~,
\end{eqnarray}
leads to the result (see Appendix \ref{app:wick})
\begin{eqnarray}
\label{b2}
\frac{1}{2}\int {\rm d}^3 x ~: B_i^{a2}({\bf  A}): &=&  g^2\frac{3}{2}
\frac{N_c}{V} (N_c^2 - 1 ) {(\tilde{C})}^2
+\frac{1}{2} \sum_{p\neq 0} \left[(p^2 + 2 g^2 N_c \tilde{C}/V) \delta_{ij}
 - p_i
p_j \right] : A_i^a(p) A_j^a(-p) :
\nonumber\\&&+\frac{1}{4 V} g^2 \sum_{p_1 \dots p_4\neq 0}f^{abc} f^{ade}
:A_j^b(p_1)A_k^c(p_2)A_j^d(p_3)A_k^e(p_4) :
\delta_{p_1+p_2+p_3+p_4,0} ~.
\end{eqnarray}
The first term corresponds to the conventional definition
of the gluon condensate \cite{zakharov}
\begin{eqnarray}
\label{26}
G^2 &=& \frac{ g^2}{\pi^2} < \frac{1}{4} F^a_{\mu\nu} F^{\mu\nu a} > =
\frac{g^2}{2 \pi^2} < B^2_i >  \\ \nonumber
&=& \frac{N^2_c - 1}{2\pi^2 3N_c} (3N_c  g^2 \tilde{C}/V)^2~.
\end{eqnarray}
The second term in Eq.(\ref{b2})  includes the mass of the quasigluons
which have both
transverse and longitudinal parts, as the operation of the
reordering destroys the projection properties of the non-Abelian
magnetic field. This fact can be understood as
spontaneous gauge symmetry breaking, and is the Bose-analogy
of the spontaneous chiral symmetry breaking  for the constituent quarks
\cite{ad,lopr,pk2s}, which is also realized by the corresponding Bogoliubov
transformation  of the ''squeezed'' type. It is well known that a
massive vector field requires for the description a number of degrees
of freedom which exceeds that provided after fixation of a gauge in QCD.
Note that the fixing of a gauge results in an elimination of the
longitudinal components of the vector field and is inconsistent with the
concept of a mass \cite{slavnov}. So, the  method of projection onto gauge
invariant variables \cite{dirac} used here is more adequate to the
phenomenon of spontaneous gauge symmetry breaking
than the conventional gauge fixing method.
Similarly to the above magnetic energy the projector $\Pi_{ij}^{ab}$ in the
electric energy $E_i^a \Pi_{ij}^{ab}({\bf A}) E_j^b$ in (\ref{QCDHAM})
is destroyed by the Wick reordering procedure leading to the
kinetic energy contribution
\begin{eqnarray}
\label{ekin}
:E_i^a   E_j^b : <\Pi_{ij}^{ab}({\bf  A})>  ~.
\end{eqnarray}
The expression $<\Pi_{ij}^{ab} ({\bf  A})>$
can be determined
in the low energy limit $ p^2 \sim 0 $:
\begin{eqnarray}
\label{cpr}
<\Pi_{ij}^{ab}({\bf  A})>
\big|_{p \sim 0}\simeq \delta_{ij}\delta^{ab} - <f^{agc} A^g_i
\frac{1}{\sum_k f^{cme} A^m_k f^{end} A^n_k}f^{dlb} A^l_j >
= \frac{2}{3} \delta_{ij}\delta^{ab} ~.
\end{eqnarray}
which can easily be checked by summation over colour and space indices.
The Eqs. (\ref{b2}), (\ref{ekin}) and (\ref{cpr}) allow us to find the
effective
Hamiltonian for the quasigluon excitations in the low-energy limit:
\begin{eqnarray}
\label{lowHam}
H_{\rm eff} ({\bf E},{\bf  A}) = \frac{1}{2} (\frac{2}{3})
 \int {\rm d}^3 x (E_i^2 + m_g^2 {A}_i^2 )~,
\end{eqnarray}
with the quasigluon mass
\begin{equation}
\label{mg2}
m_g^2 = 3 N_c g^2 \frac{\tilde{C}}{V}~.
\end{equation}
The corresponding effective low energy Lagrangian is
\begin{eqnarray}
\label{lquglu}
{\cal  L}_{\rm eff} ({\bf  A}) =
 \frac{1}{2} (\frac{2}{3}) \int {\rm d}^3 x (\dot{A}_i^2
- - m_g^2 {A}_i^2 )~.
\end{eqnarray}
The quasigluon mass $m_g$ in the low energy limit is determined
by the vacuum expectation value using the relations (\ref{26})
and (\ref{mg2}):
\begin{eqnarray}
\label{mg}
m_g = \sqrt{\frac{3\pi G}{2}}~.
\end{eqnarray}
We have shown that the Bogoliubov condensation of gluon pairs leads to a
nonvanishing contraction $\tilde{C}$ of gluon fields
which results in the spontaneous
gauge symmetry breaking and the occurence of a gluon mass.
We have obtained a new gluon mass formula for the low energy limit of QCD.

In the following Section we examine consequences of the present approach to
the low energy sector of QCD for the $\eta'$ mass formula.

\section{Applications: Quasigluon mass and $\eta -\eta'$ mass difference }
\label{sec:app}

According to the Bogoliubov condensate approach, the contraction
$\tilde{C}$ is a phenomenological parameter of the "squeezed"
vacuum state and is
directly related to the macroscopic occupation of the zero momentum
quasigluon state and therefore to the gluon mass.
We suggest to use this relationship for the fixation of
the squared mass difference
\begin{eqnarray}
\label{eta}
m_{\eta'}^2 - m_{\eta}^2 = \Delta {m^2_{\eta'}} = 0.616\; {\rm GeV}^2.
\end{eqnarray}
According to conventional approaches to the determination of the $\eta'$ mass
\cite{rst},
we suppose that the mass difference (\ref{eta}) is determined by
the $\eta' \rightarrow \eta'$ transition
through the process of the anomalous decay of the $\eta'$ meson
into the gluon condensate $B_i^a$ and a collective gluon excitation
$E^a_i$.
The effective Lagrangian of such a process can be derived according to Ref.
\cite{rst}, see also \cite{volkov},
with the result
\begin{eqnarray}
\label{lag}
{{\cal  L}_{\eta'}} = \frac{1}{4} F^a_{\mu\nu}\tilde{F}^{a\mu\nu}~\eta' c_\pi~,
\end{eqnarray}
where $c_\pi=\sqrt3 \alpha_s /({\pi F_{\pi}})$,
$\alpha_s = g^2/4\pi$ and $F_\pi = 93$ MeV.

For the derivation of an effective Hamiltonian for this process, we use
the sum of the Lagrangians (\ref{lag}) and  (\ref{lquglu}),
\begin{eqnarray}
\label{ltot}
{\cal  L}_{\rm eff}(A)={1 \over 2}(\dot{A}_i)^2- c_\pi \eta'\dot{A}_i B_i
  + \dots~.
\end{eqnarray}
 From (\ref{ltot}) follows
\begin{eqnarray}
E_i=\dot{A}_i - c_\pi \eta'B_i~,
\end{eqnarray}
such that the effective Hamiltonian reads
\begin{eqnarray}
{\cal  H}_{\rm eff}&=& E_i\dot{A}_i - {\cal  L}_{\rm eff}(A)\noindent \\
&=& \frac{1}{2} (E_i + c_\pi \eta'B_i)^2 + \dots~.
\end{eqnarray}
The effective Hamiltonian of the $\eta' \rightarrow \eta'$ transition
has the form
\begin{eqnarray}
{{\cal  H}}_{\eta'\rightarrow \eta'}=
{1 \over 2}(\eta')^2 (B_i^{a}(b))^2 c_\pi^2~,
\end{eqnarray}
where  $b$ is the constant part of the $A$ field.

Thus, we have
\begin{eqnarray}
\label{eta2}
\Delta {m^2_{\eta'}} = (\frac{3\alpha_s^2}{\pi^2 F_{\pi}^2})~<
(B_i^a(b))^{2} >.
\end{eqnarray}
This equation together with Eqs.(\ref{26}) and (\ref{mg}) leads to a relation
between the quasigluon and $\eta'$ masses:
\begin{eqnarray}
\label{eta3}
\Delta {m^2_{\eta'}} =\frac{2}{3} \frac{\alpha_s}{\pi^3 F_{\pi}^2}~m_g^4~.
\end{eqnarray}
Choosing  in the low-energy region $\alpha_s \simeq 1 $ we can estimate
>from this formula the value of the quasigluon mass and by formula (\ref{mg})
then the corresponding value of the gluon condensate as
\begin{eqnarray}
\label{mg1}
m_g = 0.71~ {\rm GeV}~~,
\ \ \ \ G^2 = 0.011 ~{\rm GeV}^4~,
\end{eqnarray}
which is in the agreement with earlier estimates \cite{biro1,zakharov}.
Note that the process of the decay of $\eta'$ into gluon fields
by means of the Hamiltonian (\ref{lag}) is forbidden, as
the vacuum expectation
value from the magnetic field $< 0_B|B_i^a(b)|0_B >$ is equal to zero.

\section{Conclusions}
\label{sec:end}

In the present paper we have considered the consequences of a squeezed
vacuum for the single particle excitation spectrum in the gluon sector of QCD
by applying the concept of the Bogoliubov theory of superfluidity to field
theory.
We have considered a squeezed homogeneous colourless condensate of zero
momentum gluon pairs.
The macroscopic occupation (squeezing) of the zero momentum mode has been
achieved through Wick reordering of the QCD Hamiltonian and is
characterized by a parameter $\tilde{C}$ which describes the magnitude of the
condensate.

The presence of the condensate leads to the occurrence of a gluon mass
and thus to spontaneous gauge symmetry breaking, i.e.
the gauge invariance of the Hamiltonian is not shared by the squeezed vacuum.
Instead of eliminating unphysical degrees of freedom by fixing a gauge
we use a projection operator method
resulting from the formal solution of Gauss law.
We show that the occurence of a condensate leads to a destruction of the
projection property
so that the generation of a mass is accompanied by the appearance
of the necessary longitudinal component for the gauge field.
We found that the quasigluon
spectrum depends on the parameter $\tilde{C}$
of the squeezed representation,
which yielded a relation between the quasigluon mass and the
gluon condensate.
We have fixed the quasigluon mass from the squared mass difference
$m_{\eta'}^2 - m_{\eta}^2 = 0.616 ~ {\rm GeV}^2$ of $\eta$ and
$\eta'$ and found that the corresponding value of the gluon condensate
$G^2 = 0.012 {\rm GeV}^4$ then agrees well with the
standard value $G^2 = 0.01 {\rm GeV}^4$ obtained by
Shifman, Vainshtein and Zakharov.

In this paper we have
populated the zero momentum state directly with quasigluons whose
dispersion relation was then determined selfconsistently by
demanding diagonality of the one-particle sector of the
Hamiltonian for nonzero momentum.
This should be considered as a first attempt to explain the concept
of the squeezed vacuum.
In a more rigorous treatment the zero momentum state should first be occupied
macroscopically with massless gluons using the freedom due to the
infrared singularity of massless theories, and  the resulting
one-particle sector of the Hamiltonian should then be diagonalized
by transformation to quasiparticles.
This has been carried out for the much simpler $\lambda\phi^4$ theory
in a separate work \cite{PaBla} which shows how in a more rigorous treatment
the renormalization of both the mass and the bare coupling are included.

Important  extensions of the present approach include the study of
small deviations from a homogeneous condensate, the inclusion of quark degrees
of freedom and the generalization to finite temperatures.
These issues are currently under investigation and will be reported in a
forthcoming paper.

\section*{Acknowledgement}
We are grateful to D. Ebert, A.V. Efremov, E.A. Kuraev, H. Leutwyler,
L.N. Lipatov, D.V. Shirkov, A.A. Slavnov,
O.I. Zav'ialov for fruitful discussions.
We thank  S. Schmidt and D. Rischke for useful comments and
critical reading of the manuscript.
The work was supported in part by the RFFI, Grant No. 95-02-14411 and
the Federal Minister for Research and Technology (BMFT) within
Heisenberg-Landau Programme.
Two of us (V.N.P. and M.K.V.) acknowledge the financial support provided
by the Max-Planck-Gesellschaft and the hospitality of the MPG Arbeitsgruppe
''Theoretische Vielteilchenphysik'' at the Rostock University, where part
of this work has been done.
One of the authors (V.N.P.) acknowledges the hospitality of the
International Centre for Theoretical Physics in Trieste.

\begin{appendix}
\section{The weakly nonideal Bose gas model}
\label{app:bogmod}

The Bogoliubov theory of the weakly interacting Bose gas \cite{nn} is
described by the nonrelativistic Hamiltonian
\begin{eqnarray}
\label{w}
H = \sum_p \frac{p^2}{2m}\,a^+_p a_p +
\frac{U_0}{2V}\,\sum_{p_1 p_2 p_1^{'} p_2^{'}}\,
a^+_{p_1} a^+_{p_2} a_{p^{'}_2} a_{p^{'}_1}
\delta_{p_1+p_2, p_1^{'}+ p_2^{'}}~.
\end{eqnarray}
The operators $a^+_p, a_p$ are the creation and annihilation operators
of bosons in the state $p$ satisfy the commutation relations
\begin{eqnarray}
\label{cr}
[a_p,a^+_{p^{'}}]= \delta_{{pp}^{'}},
\end{eqnarray}
where $p$ stands for momentum and internal quantum numbers.
The coupling constant $U_0$ is defined by the scattering amplitude of
slow particles,  $V$ is the volume of the system. We figure out the original
work of Bogoliubov \cite{nn},
for a more recent presentation of the theory of the
weakly interacting Bose gas see \cite{FW}, \cite{R}.

In the  Bogoliubov derivation  of the superfluid
spectrum  one can distinguish three points:

\begin{enumerate}
\item A macroscopic occupation of the zero momentum
state ($p=0$) is assumed so that in the thermodynamic limit a finite density
of the condensate
\begin{eqnarray}
\label{nb}
n_B = {\mbox{\rm lim}}_{\rm th} \frac{N_0}{V} \neq 0
\end{eqnarray}
occurs, where  $N_0$  denotes the number of particles in the condensate.
Therefore, the operators $a^+_0,a_0$ in the thermodynamic limit (\ref{nb})
are described as c-numbers
\begin{eqnarray}
\label{N}
a_0  \simeq a^+_0 \simeq \sqrt{N_0}.
\end{eqnarray}
This description, strictly speaking, should be completed by defining a
representation for the condensate which in the present work
is given below.

\item The next step is the expansion of the Hamiltonian (\ref{w})
around these c-numbers
\begin{eqnarray}
\label{exp}
\sum a^+_{p_1} a^+_{p_2} a_{p^{'}_2} a_{p^{'}_1} & = &
N^2_0 + N_0 \sum_{p \neq 0} (a^+_p a^+_{-p} + a_p a_{-p} + 4a^+_p a_p) +
O[a_{p\neq 0}^3]  ~.
\end{eqnarray}

Taking into account the conservation of the total number of particles $N$ and
rewriting
$$
N_0=N - \sum_{p\neq 0}a^+_p a_p ~~~;
\left (\frac{N-N_0}{N} \ll 1
\right ),
$$
in Eq. (\ref{exp}) and neglecting terms of higher than second order
in the particle operators $a_{p\neq 0}, a^+_{p\neq 0} $,
the Hamiltonian (\ref{w}) transforms into
\begin{eqnarray}
\label{ha2}
H =\frac{N}{2} \nu + \sum_{p\neq 0} \left [a^+_p a_p \varepsilon _p +
\frac{\nu}{2}\;(a^+_p a^+_{-p} + a_p a_{-p}) \right ] + O[a_{p\neq 0}^3],
\end{eqnarray}
where
\begin{eqnarray}
\varepsilon_p =\frac{p^2}{2m} + \nu,  ~~~ \nu=U_0\frac{N}{V}.
\end{eqnarray}

\item The last step is the diagonalization of (\ref{ha2}) using the Bogoliubov
transformation, i.e. the transition to the operators of quasiparticles $b^+_p $
and $b_p $ for $p \neq 0$
\begin{eqnarray}
\label{b1}
b_p &=& U^{-1} a_p U = {\rm cosh}(f_p) a_p + {\rm sinh}(f_p) a^+_{-p},
\nonumber\\
b^+_p&=& U^{-1} a^+_p U = {\rm cosh}(f_p) a^+_p + {\rm sinh}(f_p) a_{-p},
\\
\nonumber
\end{eqnarray}
where
\begin{eqnarray}
\label{b}
U =\exp \left \{\sum_p \frac{f_p}{2} (a^+_pa^+_{-p}\, -
\,a_p a_{-p}) \right \}~.
\end{eqnarray}
The $b_p$ satisfy the same commutation relations as the $a_p$.
The function $f _p  $ is found from the requirement of the disappearance
of nondiagonal terms as
\begin{eqnarray}
\label{fp}
f_p = \frac{1}{2} {\rm arth} \left[\frac{\nu}{\varepsilon_p}\right]~,
\end{eqnarray}
so that the Hamiltonian (\ref{ha2}) gets the form
\begin{eqnarray}
H =\frac{N}{2} \nu -{1\over 2}\sum_{p\neq 0}
\left(\varepsilon_p-\omega_B(p)\right)
+ ~ \sum_{p\neq 0} b^+_p b_p~ \omega_B(p) +~ O[b_{p\neq 0}^3]~,
\end{eqnarray}

where $\omega_B$ is the spectrum of excitations in a superfluid liquid
\begin{eqnarray}
\label{bf}
\omega^2_B(p) = \varepsilon^2_p - \nu^2 =\left (\frac{p^2}{2m}\right )^2 +
\frac{p^2}{2m} \;\left ( 2U_0\; \frac{N}{V} \right ) \;\;,
\end{eqnarray}
which is determined by the condensate density
$n_B = N_0/V \cong N/V$ and by the coupling constant
$U_0$.
\end{enumerate}
In the low momentum region this expression describes  the Landau sound
and the particle excitations with energy $\left (p^2/2m\right)$
disappear.

Note that the vacuum energy $E_0$ contains a divergent sum which can be
renormalized by expressing it in terms of the physical scattering length $a$
instead of the bare coupling $U_0$ (see \cite{FW}, p. 318).

In his paper \cite{nn}, Bogoliubov did not determine the
representation of the condensate state for which
Eq. (\ref{N}) is fulfilled.
Usually one  assumes that of the coherent state
\begin{eqnarray}
\label{c}
|0_{\rm C} >~ = \exp\left\{\sum_p c_0(a^+_0 + a_0)\right\}~ |0>~,~
c_0 =\sqrt{N_0}~,
\end{eqnarray}
for which holds
\begin{eqnarray}
<0_{\rm C}|a_0|0_{\rm C}>~=~<0_{\rm C}|a^+_0|0_{\rm C}>~=\sqrt{N_0}~,
\end{eqnarray}
corresponding to Eq. (\ref{N}).

However, to get the Bogoliubov result it is enough to assume
the weaker condition
\begin{eqnarray}
\label{nn}
(a^+_0)^2 \simeq a^2_0 \simeq a^+_0\; a_0 \sim N_0~,
\end{eqnarray}
rather than (\ref{N}).
These relations are fulfilled for the representation of the condensate state
which is given by the same Bogoliubov transformation as for
$p \neq 0$ (\ref{b}):
\begin{eqnarray}
\label{s}
|\,0_{B} > = U_{B}^{-1}~~|~0 >~,
\end{eqnarray}
where
\begin{eqnarray}
\label{b0}
U_{B_0} =\exp \left \{ \frac{f_0}{2} (a^+_0a^+_{-0}\, - \,a_0 a_{-0}) \right
\}~.
\end{eqnarray}

The inverse of the unitary operator (\ref{b}) defines also
the transformation of the old into a new vacuum state for momenta $p \neq 0$.
In quantum optics the vacuum $ |\,0_B >$  is called 'squeezed vacuum', see e.g.
\cite{optics}.

For the ''squeezed'' vacuum representation
(\ref{s}) of the condensate we have  the realization (\ref{nn})
\begin{eqnarray}
\label{a2}
<0_B\,|\,a^2_0\,|\,0_B>
& = & <0_B\,|\,(a^+_0)^2\,|\,0_B > ~= - {\rm cosh}{f_0} ~{\rm sinh}{f_0} \;\;,
\\
\nonumber
<0_B\,|\,a^+_0 a_0\,|\,0_B> & = & ({\rm sinh}{f_0})^2 = N_0~,
\label{a+a}
\end{eqnarray}
and at large $N_0$ (\ref{nn}) means that
\begin{eqnarray}
\label{n0}
- -{\rm cosh}{f_0}~ {\rm sinh}{f_0} &\simeq & ({\rm sinh}{f_0})^2
\simeq N_0 \rightarrow \infty\;\;, \nonumber\\
f_0 \sim -\frac{1}{2}\ln{4N_0}~.
\end{eqnarray}
The choice of the squeezed vacuum is more prefable from the point view
of a general consideration of all momenta, $p=0$ {\it and} $p \neq 0$.
Together, the Bogoliubov transformation now is given by the product $UU_{B}$.

\section{Gauge invariant variables}
\label{app:gaugi}

The unphysical components of the gluon fields are formally eliminated by the
transformation to gauge invariant variables \cite{dirac,perv1} which are
functionals constructed using the solution (\ref {10})
\begin{eqnarray}
\label{12}
{A}^I_i [{\bf  A}] \equiv V({\bf  A}) ({A}_i + \partial_i) V({\bf  A})^{-1}~.
\end{eqnarray}
The matrix $V$ is defined from the equation
\begin{eqnarray}
\label{DefV}
V({A}_0[{\bf  A}] + \partial_0) V^{-1} = 0 \Rightarrow V({\bf  A}) =
 T{\rm exp}(\int^t
{A}_0[{\bf  A}] dt^{'})
\end{eqnarray}
(up to a stationary matrix as the time boundary condition).
The invariance of these functionals under arbitrary time dependent
gauge transformations $v({\bf  x},t)$
\begin{equation}
\label{GI2}
{A}^I_i [{\bf  A}^v] = V({\bf  A}) v^{-1} v ({A_i} + \partial_i)
v\ v^{-1} V({\bf  A}) = {A}^I_i [{\bf  A}]~,
\end{equation}
follows from the transformation properties of $A_i$ in (\ref{GI})
and of $V({\bf  A})$:
\begin{eqnarray}
V({\bf  A}^v) = V({\bf  A}) v^{-1}~.
\end{eqnarray}
which follows from (\ref{DefV}) and (\ref{GI}).
As consequence of this the variables (\ref{12}) represent only $2~(N_c^2-1) $
independent degrees of freedom. They contain hidden projection operators
onto  generalized transverse components similar to the magnetic field which
satisfies the Bianchi identity (\ref{BI}).
The projection operator is contained (different to the QED case) not in the
$A_i^I$ themselves, but only in their time derivatives
$\dot{A}_i^I$ which satisfy the ``Bianchi type'' identities
\begin{equation}
\label{DAI}
D^{ab}_i({\bf  A}^I) \dot{A}_i^{I^b} \equiv 0~.
\end{equation}
In the terms of the functionals (\ref{12}) the Lagrangian (\ref{1})
takes the form
\begin{equation}
\label{17}
{\cal  L}^{Red} ({\bf  A}^I) = \frac{1}{2}~ \left [\dot{A}^{I^a2}_i
- - B^{a2}_i({\bf  A}^I) \right ]~.
\end{equation}
The canonical momenta to the spatial fields $A^{Ia}_i$
are
\begin{equation}
{E^I_i}^a \equiv\frac{\delta {\cal  L}}{\delta \dot{A^I_i}^a}= \dot{A^I_i}^a~.
\end{equation}
The Hamiltonian becomes
\begin{equation}
H^{Red}({\bf  A}^I,{\bf  E}^I)=
\int d^3x {1\over 2}\left[E_i^{Ia2}+B_i^a({\bf  A}^I)^2  \right]~.
\end{equation}
It follows from (\ref{DAI}) that the electric fields ${E^I}_i^a $ satisfy
the Gauss constraint
\begin{equation}
D^{ab}_i({\bf  A}^I) {E^I}^b_i = 0~.
\end{equation}
Like the Bianci identity (\ref{BI}) for the magnetic field, this shows
the generalized transversality of the invariant
electric fields ${E^I}^a_i$.

In order to quantize the theory one could then like in QED
impose the following canonical commutation relations on the physical
variables $A^I$ and $E^I$:
\begin{equation}
[E_i^{Ia}({\bf  x}),A_j^{Ib}({\bf  x}')]=
i\delta^{ab}\delta_{ij}\delta({\bf  x}-{\bf  x}')~.
\end{equation}
In QCD, however, this leads to a contradiction when applying the
covariant derivative on it.

Instead one has to impose canonical commutation relations directly on
the three cartesian fields $A_i$ and $E_i$ and write both $E^I$ and $A^I$
as functionals of $E$ and $A$. Whereas the form of $A^I[{\bf  A}]$ is known
by construction (\ref{12}), the functional form of $E^I[{\bf  E},{\bf  A}]$
 in terms of
$E$ and $A$ has not been found yet, but is subject of intensive research
\cite{Kved}.
However, even if this problem is solved there remains still the question of
correct ordering of the operators $A$ and $E$.

\section{Wick reordering of gluon fields}
\label{app:wick}
In this appendix we perform the Wick reordering of the magnetic part of the
Hamiltonian (\ref{17}),
\begin{eqnarray}
\label{b2n}
\frac{1}{2}\int {\rm d}^3 x ~:{B_i^a}^2({\bf  A}): &=&
\frac{1}{2}\int {\rm d}^3 x ~\left\{:(\partial_j A_k^a) \delta^T_{kl}
(\partial_jA_l^a):
+2 g f^{abc} :(\partial_j A_k^a) A_j^b A_k^c:
+\frac{1}{2} g^2 f^{abc} f^{ade} : A_j^bA_k^cA_j^dA_k^e : \right\}~,
\end{eqnarray}
which reads in momentum space
\begin{eqnarray}
\label{b2p}
\frac{1}{2}\int {\rm d}^3 x ~ : {B_i^a}^2({\bf  A}) : &=&
\frac{1}{2}\bigg\{\sum_{p} : A_k^a(p)\left[p_kp_l - p^2
\delta_{kl}\right]A_l^a(-p) :\nonumber\\
& &+2 g^2 \frac{1}{\sqrt{V}} f^{abc} \sum_{p_1 p_2} i(p_1+p_2)_j :
A_k^a (p_1+p_2) A_j^b(-p_1) A_k^c(-p_2) :\nonumber\\
& & + \frac{1}{2}  g^2 \frac{1}{V}f^{abc} f^{ade}
\sum_{p_1 \dots p_4} :A_j^b(p_1)A_k^c(p_2) A_j^d(p_3) A_k^e(p_4) :
\delta_{p_1+p_2+p_3+p_4,0}\bigg\}~.
\end{eqnarray}
The reodering of the first two terms on the r.h.s. of Eq. (\ref{b2p}) gives
no extra contraction contribution.
For the Wick reordering of the third term we write
\begin{eqnarray}
\label{a4}
f^{abc} f^{ade} : A_j^b(p_1)A_k^c(p_2)A_j^d(p_3)A_k^e(p_4) : &=&
f^{abc} f^{ade} A_j^b(p_1)A_k^c(p_2)A_j^d(p_3)A_k^e(p_4)\nonumber\\
&&-f^{abc} f^{ade} <A_j^b (p_1)A_j^d(p_3)> A_k^c(p_2) A_k^e(p_4)\nonumber\\
&&
- -f^{abc} f^{ade} A_j^b(p_1) A_j^d(p_3) <A_k^c(p_2) A_k^e(p_4)>\nonumber\\
&&-f^{abc} f^{ade}   <A_j^b(p_1) A_k^c(p_2)> A_j^d(p_3) A_k^e(p_4)\nonumber\\
&&
- -f^{abc} f^{ade}  A_j^b(p_1) A_k^e(p_4) <A_k^c(p_2)  A_j^d(p_3)> \nonumber\\
&&+f^{abc} f^{ade} <A_j^b(p_1) A_j^d(p_3)> <A_k^c(p_2) A_k^e(p_4)>\nonumber\\
&&
+f^{abc} f^{ade}   <A_j^b(p_1) A_k^c(p_2)> <A_j^d(p_3) A_k^e(p_4)>~.
\end{eqnarray}
Using formula (\ref{aa}) and the identities
\begin{eqnarray}
\sum_{abc} f^{abc} ~f^{abc} &=& N_c (N_c^2 -1)~,\nonumber\\
\sum_{ab} f^{abc} ~f^{abe} &=& N_c \delta_{ce}~,\nonumber
\end{eqnarray}
in Eq. (\ref{a4}) we thus have
\begin{eqnarray}
\label{25}
\sum_{p_1 \dots p_4}f^{abc} f^{ade} &:
& A_j^b(p_1)A_k^c(p_2)A_j^d(p_3)A_k^e(p_4) :
\delta_{p_1+p_2+p_3+p_4,0} = \nonumber\\
&&  6 N_c (N_c^2 - 1 ) C^2
- - 4 N_c C \sum_{p} A_i^a(p) A_i^a(-p) \nonumber\\
&& + \sum_{p_1 \dots p_4}f^{abc} f^{ade}
A_j^b(p_1)A_k^c(p_2)A_j^d(p_3)A_k^e(p_4)
\delta_{p_1+p_2+p_3+p_4,0}\nonumber\\
&=& 6 N_c (N_c^2 - 1 ) {(C^B)}^2
+4 N_c C^B \sum_{p} ::A_i^a(p) A_i^a(-p)::  {\rm e}^{f_p +f_{-p}}
\nonumber\\&&+ \sum_{p_1 \dots p_4}f^{abc} f^{ade}
::A_j^b(p_1)A_k^c(p_2)A_j^d(p_3)A_k^e(p_4) ::
{\rm e}^{f_{p_1} +f_{p_2} +f_{p_3} +f_{p_4} }
\delta_{p_1+p_2+p_3+p_4,0} ~.
\end{eqnarray}
Putting all together,
\begin{eqnarray}
\label{b2r}
\frac{1}{2}\int {\rm d}^3 x ~: {B_i^a}^2({\bf  A}): &=&  g^2\frac{3}{2}
\frac{N_c}{V} (N_c^2 - 1 ) {(C^B)}^2
+\frac{1}{2} \sum_{p} \left[(p^2 + 2 g^2 N_c C^B/V) \delta_{ij} - p_i
p_j \right] :: A_i^a(p) A_j^a(-p) ::  {\rm e}^{f_p +f_{-p}}
\nonumber\\&&+\frac{1}{4 V} g^2 \sum_{p_1 \dots p_4}f^{abc} f^{ade}
::A_j^b(p_1)A_k^c(p_2)A_j^d(p_3)A_k^e(p_4) ::
{\rm e}^{f_{p_1} +f_{p_2} +f_{p_3} +f_{p_4} }
\delta_{p_1+p_2+p_3+p_4,0} ~.
\end{eqnarray}
This proves the result of Eq. (\ref{b2}) used in the main text.
\end{appendix}


\newpage

\begin{thebibliography}{99}
\bibitem{savvidy}
 G.K. Savvidy: Phys. Lett. {\bf B 71} (1977) 133;\\
 S.G. Matinyan and G.K. Savvidy: Nucl. Phys. {\bf B 134} (1978) 539;\\
 N.K. Nielsen and P. Olesen: Nucl. Phys. {\bf B 144} (1978) 376
\bibitem{coleman}
 S. Coleman: Ann. Phys. {\bf 101} (1976) 239
\bibitem{leutwyler}
 H. Leutwyler: Nucl. Phys. {\bf B 179} (1981) 129
\bibitem{belavin}
 A. Belavin, A. Polyakov, A. Schwartz and Y. Tyupkin:
 Phys. Lett. {\bf 59 B} (1975) 85
\bibitem{zakharov}
 M.A. Shifman, A.I. Vainshtein and V.I. Zakharov:
 Nucl. Phys. {\bf B 147} (1979) 385, 448, 519
\bibitem{shuryak}
 E. Shuryak and M. Velkovsky: Phys. Rev. {\bf D 50} (1994) 3323
\bibitem{nn}
 N.N. Bogoliubov: J. Phys. {\bf 11} (1947) 23
\bibitem{umt}
 H. Umezawa, H. Matsumoto and M. Tachiki: {\it Thermo Field Dynamics and
 Condensed States}, North Holland, Amsterdam 1982
\bibitem{celenza}
 L.S. Celenza and C.M. Shakin: Phys. Rev. {\bf D 34} (1986) 1591
\bibitem{biro1}
 T.S. Biro: Ann. Phys. (NY) {\bf 191} (1989) 1; Phys. Lett. {\bf B 278} (1992)
 15
\bibitem{biro2}
 T.S. Biro: Int. J. Mod. Phys. {\bf 2} (1992) 39
\bibitem{mishra1}
 A. Mishra, H. Mishra, S.P. Misra and S.N. Nayak:
 Phys. Rev. {\bf D 44} (1991)
 110;
 Z. Phys. {\bf C 37} (1993) 233
\bibitem{mishra2}
 A. Mishra, H. Mishra and S.P. Misra: Z. Phys. {\bf C 57} (1993) 241;
 Z. Phys. {\bf C 59} (1993) 159
\bibitem{slavnov}
 A.A. Slavnov and L.D. Faddeev:
 Introduction to Quantum Theory of Gauge Fields,
 Nauka,  Moscow 1988
\bibitem{cw}
 S. Coleman and E. Weinberg: Phys. Rev. {\bf D 7} (1973) 1888
\bibitem{Georgi}
 M. Georgi and S.L. Glashow: Phys. Rev. Lett. {\bf 28} (1972) 1494
\bibitem{CC}
 P. Castorina and M. Consoli: Phys. Lett. {\bf 131 B} (1983) 351
\bibitem{PaBla}
 H.-P. Pavel, D. Blaschke, G. R\"opke, and V.N. Pervushin:
 {\it Bogoliubov Condensation of Pairs in massless $\lambda\phi^4$},
 in preparation
\bibitem{dirac}
 P.A.M. Dirac: Can. J. Phys. {\bf 33} (1955) 650
\bibitem{perv1}
 V.N. Pervushin: Teor. Mat. Fiz. {\bf 45} (1980) 394 (russ.);
 Sov. J. Theor. Math. Phys. {\bf 45} (1981) 1100
\bibitem{perv2}
 V.N. Pervushin: Riv. Nuovo Cimento {\bf 8} (10) (1985) 1
\bibitem{fj}
 L.D. Faddeev and R. Jackiw: Phys. Rev. Lett. {\bf 60} (1988) 1962
\bibitem{kp}
 A.M. Khvedelidze and V.N. Pervushin: Helv. Phys. Acta {\bf 67} (1994) 321
\bibitem{ad}
 S. Adler and C. Davis: Nucl. Phys. {\bf B 244} (1984) 469
\bibitem{lopr}
 A. LeYaouanc, Ll. Oliver, O. Pene and J.-C. Raynal:
 Phys. Rev. {\bf D 29} (1984) 1233, {\bf D 31} (1985) 137
\bibitem{pk2s}
 V.N. Pervushin, Yu.L. Kalinovsky, W. Kallies and N.A. Sarikov:
 Fortschr. Phys. {\bf 38} (1990) 333
\bibitem{optics}
 W. Vogel and D.-G. Welsch: {\it Quantum Optics}, Akademie- Verlag,
 Berlin, 1994
\bibitem{ChrLee}
 N.H. Christ and T.D. Lee: Phys. Rev. {\bf D 22} (1980) 939
\bibitem{RJ}
 R. Jackiw: Rev. Mod. Phys. {\bf 52} (1980) 661
\bibitem{GJ}
 J. Goldstone and R. Jackiw: Phys. Lett. {\bf 74B} (1978) 81;\\
 V. Baluni and B. Grossman: Phys. Lett. {\bf 78B} (1978) 226;\\
 Yu.A. Smirnov: Sov. J. Nucl. Phys. {\bf 41 (5)} (1985) 835
\bibitem{Lenz}
 F. Lenz, H.W.L. Naus and M. Thies: Ann. Phys. (NY) {\bf 233} (1994) 317
\bibitem{rst}
 C. Rosenzweig, J. Schechter and C.G. Trahern, Phys. Rev. {\bf D 21} (1980)
3388
\bibitem{volkov}
 M.K. Volkov: Fiz. Elem. Chastits At. Yadra {\bf 17} (1986) 433,
 Sov. J. Part. Nuclei {\bf 17} (1986) 186.
\bibitem{FW}
 A.L. Fetter and J.D. Walecka: {Quantum Theory of Many-Particle Systems},
 McGraw-Hill, New York 1971
\bibitem{R}
 G. R\"opke, Ann. Phys. (Leipzig) {\bf 3} (1994) 145
\bibitem{Kved}
 S.A. Gogilidze, A.M. Khvedelidze and V.N. Pervushin:
 {\it Abelianization of First Class Constraints}, Phys. Rev. {\bf D}, in press
\end{thebibliography}
\end{document}